# A Time-Optimized Content Creation Workflow for Remote Teaching

Sebastian Hofstätter
TU Wien
s.hofstaetter@tuwien.ac.at

Sophia Althammer
TU Wien
sophia.althammer@tuwien.ac.at

Mete Sertkan
TU Wien
mete.sertkan@tuwien.ac.at

Allan Hanbury
TU Wien
allan.hanbury@tuwien.ac.at

## ABSTRACT

We describe our workflow to create an engaging remote learning experience for a university course, while minimizing the post-production time of the educators. We make use of ubiquitous and commonly free services and platforms, so that our workflow is inclusive for all educators and provides polished experiences for students. Our learning materials provide for each lecture: 1) a recorded video, uploaded on YouTube, with exact slide timestamp indices, which enables an enhanced navigation UI; and 2) a high-quality flow-text automated transcript of the narration with proper punctuation and capitalization, improved with a student participation workflow on GitHub. All these results could be created by hand in a time consuming and costly way. However, this would generally exceed the time available for creating course materials. Our main contribution is to automate the transformation and post-production between raw narrated slides and our published materials with a custom toolchain. Furthermore, we describe our complete workflow: from content creation to transformation and distribution. Our students gave us overwhelmingly positive feedback and especially liked our use of ubiquitous platforms. The most used feature was YouTube's chapter UI enabled through our automatically generated timestamps. The majority of students, who started using the transcripts, continued to do so. Every single transcript was corrected by students, with an average word-change of 6%. We conclude with the positive feedback that our enhanced content formats are much appreciated and utilized. Important for educators is how our low overhead production workflow was sustainable throughout a busy semester.

## CCS CONCEPTS

• **Social and professional topics** → **Computing education**;

## KEYWORDS

Remote teaching, Enhanced lecture materials, Automated post-production

## 1 INTRODUCTION

We describe our experience with our content creation and dissemination workflow used in a master-level university course in the summer semester of 2021. Following COVID-19 prevention guidelines the course was held completely remotely. The constraint of remote teaching inspired us to completely revamp the way we typically disseminated materials to students. Our aim was to provide a great learning experience for our students, while at the same time minimizing our time spent on post-production tasks. We collect and evaluate feedback from our students, in order to analyze the acceptance of the dissemination format.

From a student perspective our learning materials provide for each lecture: 1) a recorded video, uploaded on YouTube, with exact slide timestamp indices, which enables an enhanced navigation UI; and 2) a high-quality flow-text automated transcript of the narration with proper punctuation and capitalization, improved with a student participation workflow on GitHub.

The first obvious question is why do we propose a new workflow, when there are plenty of options available for creating & disseminating rich lecture content? Platforms such as ClassTranscribe [2], VideoLectures[1], or SlidesLive[2] provide a very similar end-result to our approach. However, we could not use them as we operated in the following (partially self-imposed) constraints:

- We wanted to make our content accessible and discoverable for a large audience outside our university
- We had no additional budget for commercial services
- We wanted to focus our time on creating the actual content, rather than the post-production or server hosting

We did not want to setup and host any web-based platforms ourselves [2], rather utilize ubiquitous and free to use platforms best suited for our needs: YouTube and GitHub. Furthermore, we set up our workflow around a one-time mostly automatic and time-efficient transformation process.

We utilize built-in functionality of PowerPoint to record a narration per slide. We created a custom transcription program, which uses a speech recognition service (such as Azure Speech API) to transcribe the narration per slide in a flow-text format. The program also outputs timestamps & slide titles for YouTube, which map to all slide transitions in the exported video. This mapping enables YouTube's chapter UI for enhanced navigation between slides.

Our students gave us overwhelmingly positive feedback and especially liked our use of ubiquitous platforms. The most used feature was YouTube's chapter UI enabled through our automatically generated timestamps. The majority of students, who started using the transcripts, continued to do so. Every single transcript was corrected by students, with an average word-change of 6%. We conclude with the positive feedback that our enhanced content formats are much appreciated and utilized. Important for educators

---

[1] http://videolectures.net/ (Accessed 08/08/21)
[2] https://slideslive.com/features (Accessed 08/08/21)

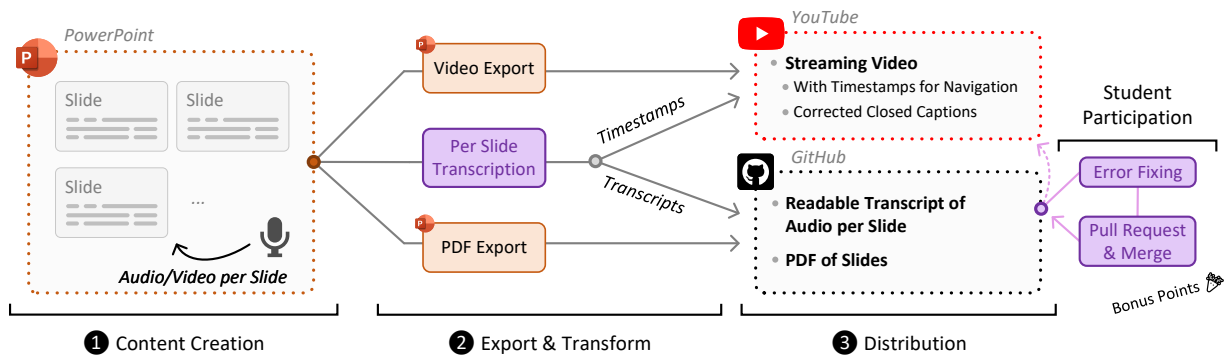

Figure 1: Overview of our content creation & dissemination workflow: ❶ We create our presentations in PowerPoint and add narrations per slide; ❷ We use both the native export functionality of PowerPoint and our custom slide transcription program; ❸ We distribute the exported formats on YouTube and GitHub. We also encourage students to fix transcripts, which we then use to improve the closed captions on YouTube as well.

is how our low overhead production workflow was sustainable throughout a busy semester. Our custom workflow tools and showcasing lecture materials are publicly available at:

*https://github.com/sebastian-hofstaetter/teaching*

### 1.1 Our Course: Information Retrieval

Our Information Retrieval (IR) course is held once a year; is featured in most master curricula as an elective or part of an elective module; furthermore it is available for interested bachelor students as extra credit. It saw a steady rise of enrolled students in the last years, with 120 registered students in the summer semester of 2021. It has a planned 75 hour total workload for students.

We provide a total of 11 pre-recorded lectures uploaded in a weekly schedule, 1 large exercise, and weekly office hours over video chat. We cover Information Retrieval basics, Natural Language Processing (NLP) basics, and dive deep into recent advancements of neural IR research. Because our students come from various academic levels and knowledge backgrounds we accompany our lecture materials with a diverse set of curated external materials.

### 1.2 Context: Information Retrieval Research

Information retrieval is the science behind search technology. Certainly the most visible instances are the large web search engines Google and Bing. Albeit, information retrieval appears everywhere we have to search unstructured data. Taking off in 2019 the information retrieval research field began an enormous paradigm shift towards utilizing BERT-based language models [10] in various forms [14, 16, 17, 23] to great effect with huge leaps in quality improvements for search results using large-scale data [3, 8, 9].

This rapid change in the research field brings the need for updated university courses. One hurdle is the lack of textbooks or simple aggregations covering the recent advances [18].[3] Another major issue we identified with teaching the machine learning advances in IR is the steep learning curve and the broad foundational knowledge required from both IR and NLP fields to effectively understand the new techniques. This brings additionally the need for condensed lectures covering the basics.[4]

## 2 CONTENT CREATION WORKFLOW

We describe our content creation & dissemination workflow – from an educators perspective as well as the student experience. Figure 1 shows a structured overview of this workflow with 3 main parts: content creation, export & transformation, and distribution.

### 2.1 Content Creation

The starting point for our workflow is a completed PowerPoint presentation. We use a single file per lecture. We do not require special treatment of the slides themselves, so one can start with existing presentations as well as creating new ones.

To record audio of our presentation, we use the native narration function of PowerPoint. The recording can be started from the first or a specific slide. It starts a slide-show like interface that allows to fluidly move to the next slide (which automatically starts a new recording on the new slide). We choose only to record audio, albeit one can use this to record a webcam video as well. We especially like the capability of stopping and restarting on every slide, as it makes breaks during the recording session, minor fixes, or adding/moving/hiding slides after the initial session very convenient.

Each recording per slide is attached to the slide via an autoplaying audio file. Additionally PowerPoint automatically sets the slide duration to the length of the recording, and remembers manually triggered animation start times. Together these information points allow PowerPoint to create a video from the slide show that looks like it was recorded via a screengrab of a live presentation. Furthermore, this file format also allows us to access the recordings, as presented in the next section.

---

[3]For example: https://twitter.com/jobergum/status/1396520526118539271

[4]For example: https://twitter.com/srchvrs/status/1362634045419646976

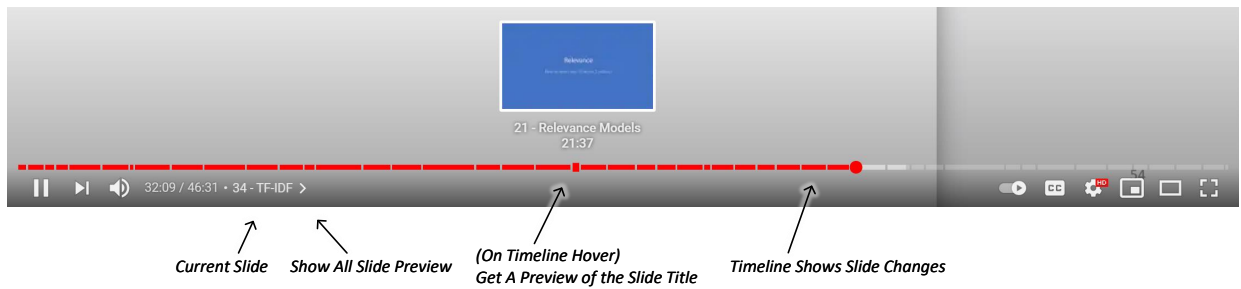

Figure 2: The User interaction modes of the YouTube chapter feature (May 2021) for video timeline progress. Enabled by our automatically generated timestamps & chapter names from our transcription procedure.

## 2.2 Export & Transformation

While we can use the native functionality of PowerPoint to create video and PDF output, we needed to create a custom program to extract the audio per slide and transcribe it. We created a python program to read any existing PowerPoint file and output transcriptions and timestamps. In essence any modern Office file (such as PowerPoint's .pptx) is just a compressed ZIP file containing various XML files and raw media assets. Therefore it is easy to open and read these files programmatically. Our python program only needs as input the filename and an API key for the Azure speech recognition API. We choose the Azure service because it provides 5 hours of free transcription per month (June 2021), which was enough for us to not pay anything for the whole course.

For every slide, we extract the audio file, convert it into Azure's input format and send it to Azure for speech recognition. The result is then written in Markdown format. Furthermore, we keep track of the length of the narration on each slide and extract the title and text on each slide. We extract the text content not to output it, but because the Azure speech API allows us to provide hints of possible appearing terms in the narration as text. Therefore we provide individual words to the Azure API as a suggestion of possible content. The extracted slide numbers and title, together with the duration information, is transformed into YouTube's timestamp text format and output as well. The timestamp format looks like minutes:seconds text, which is very simple – but creating 40-60 exact timestamps per lecture per hand is very time consuming.

In total we produce the following files and materials per lecture:

- **Slides** in PDF format (using the native PowerPoint export)
- **Video** in MP4 format (in 1080p resolution using the native PowerPoint export)
- **Transcript** in Markdown format
- **Timestamps** with slide number and title in plain text format

## 2.3 Material Distribution

As shown in Figure 1 our two main platforms for publishing content are YouTube for videos and GitHub for slides and transcripts. In this section we go into detail on how we and our students can use these platforms in an optimal way.

***YouTube.*** We uploaded the exported video to YouTube and copied the exported timestamp text into the video description. YouTube converts timestamps into user interface improvements in the video player across devices.[5] The main improvements that are enabled are highlighted in Figure 2. The video player shows the current slide name; a button to show a preview of all slides; a preview of the slide number and title (while hovering over the timeline progress bar); as well as visual breaks in the timeline progress bar for every slide transition.

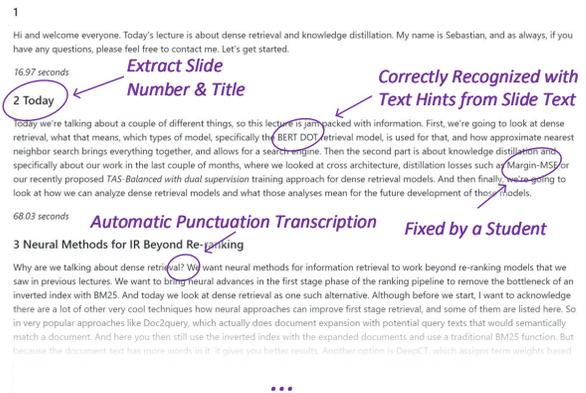

Figure 3: Visual overview of the transcript in Markdown format on GitHub with interesting sections indicated.

These improvements move the YouTube player closer to specialized slide-show players such as SlidesLive[6], VideoKen[7], or VideoLectures[8], but with the benefit of being free, offering long-term storage, and polished user interfaces with support for watching the lecture videos with chapters on all devices and browsers. Furthermore, by using YouTube's recommendations we can easily reach a larger audience beyond our students and the students get our and other's follow up videos recommended.

Finally, we created a playlist and continuously added the videos to further enhance our course discover-ability and watch-ability.

***GitHub.*** We use a single GitHub repository, first as a landing page for our course, where we introduce the contents and link to all available materials per lecture, and second as a public hosting platform for our PDF slides and transcript files. The PDF slides are read-only and not meant to be changed by the students. However, the text-file based transcripts are meant to be editable, while still providing a comfortable reading experience. For this the Markdown file format is well-suited and integrates nicely with the GitHub web user interface. Markdown is a simple text formatting language, which allows to define headings, lists, simple formatting, and tables with sets of special characters (mainly *, #, -), therefore text is readable both in the editor-view as well as the rendered view with styling. The GitHub UI automatically renders Markdown files to good looking HTML.

---

[5]Reference: https://support.google.com/youtube/answer/9884579?hl=en (Accessed: 05/19/21)
[6]Available at: https://slideslive.com/
[7]Available at: https://videoken.com/
[8]Available at: http://videolectures.net/

In Figure 3 we show an example of the reading experience in the GitHub UI for our transcripts. For each slide we output the slide number and title as section heading, followed by the transcribed text and narration time. This makes navigating the transcript straightforward and allows students to quickly lookup certain slides. The Azure speech recognition service we used also produces good punctuation and capitalization for the text.

### 2.4 Student Participation

For each new lecture we added, we invited students to use the established GitHub fork and pull request workflow to make edits to the automatic transcripts. To encourage participation we awarded bonus points for each fixed lecture. Once a lecture transcript was corrected, we used the final transcripts to update the closed captions of our videos on YouTube, as YouTube automatically maps the text back to the audio.

A problem we quickly encountered were parallel edits to the transcript, from 2 or more students who submitted different fixes to the same file. While git is meant to compensate for this with merges and version tracking, we found it is not feasible to merge these plaintext edits in a short amount of time. Therefore, we adapted our procedure and created a single GitHub issue per lecture, which students could use to block a file, edit it, and finally we could cleanly merge the changes to the main branch.

## 3 DATA ANALYSIS & FEEDBACK

In this section we present an analysis of the transcription quality, as measured by student corrections, as well as a thorough look at the feedback from our students for our content dissemination workflow and the provided material formats.

### 3.1 Automatic Transcription Quality

For every single lecture transcript we published, there was at least one student who contributed fixes to the text. This by itself is already a great result, showing both the interest of the students and the viability of the workflow using GitHub pull requests to fix errors. Now, that we have both our automatic transcripts and human corrected versions, we can use the differences to measure the initial automatic speech recognition quality.

*Methodology.* We compare the automatic with the final version of each lecture transcript using the `wdiff` utility. We count both alphanumeric character changes, as well as punctuation changes. We used the Azure speech recognition service from March to June 2021. The online service might have changed in between lectures

Table 1: Analysis of the transcript quality per lecture – measured by the word changes made by our students.

| Lecture | Words | Unchanged | Most Common Changes |
|---|---|---|---|
| 0 | 2,517 | 96% | re-ranking; So; representation; course |
| 1 | 5,172 | 94% | TF-IDF; BM25; a; and; is; we; relevance |
| 2 | 4,423 | 91% | nDCG; So; then; precision; the; have |
| 3 | 6,004 | 92% | now; a; and; MSMARCO-Passage |
| 4 | 4,915 | 91% | Word2Vec; pre-trained; that; and; word |
| 5 | 6,012 | 99% | Hadamard product; because; N-grams |
| 6 | 5,111 | 93% | pre-trained; BERT; it; and; pre-training |
| 7 | 6,641 | 92% | non-relevant; re-ranking; a; and; layer |
| 8 | 5,540 | 93% | BERT; re-ranking; a; passages; re-rank |
| 9 | 5,310 | 92% | the; and; re-ranking; BERT; of; of the |
| 10 | 6,963 | 96% | BERT DOT; BM25; BERT; re-ranking |
| **Avg.** | **5,328** | **94%** | |

and since we used it. However, we do not observe clear trends in the accuracy.

*Results.* We aggregated the transcription length in words, the ratio of unchanged (therefore correct words), as well as a selection of the most frequent corrected words per lecture in Table 1. On average the speech recognition service detected 5,328 words (this also includes the added slide titles and length description). Our students left 94% of the words untouched and corrected on average 6% of the words including punctuation fixes. Furthermore, we observe the most common changed & corrected words to be often technical acronyms and hyphenated words. This is not surprising, given that many of these acronyms are pronounced differently than the raw characters would suggest. For example, the popular acronym *BERT* is often confused with *bird*.

In conclusion, automatic speech recognition is sufficient to produce transcripts of technical jargon (with a word and punctuation error rate of 6%). Combined with our human-in-the-loop workflow we produced high quality transcripts, which contain correct acronyms & punctuation, and are arranged by slide.

### 3.2 Student Feedback

To evaluate our teaching approach, we asked our students to give us feedback via our Moodle platform after the final exam. Although technically part of the exam via a link to the feedback module, we made it very clear that the feedback is absolutely not connected to their grade. We recognize that this might bias results, albeit we ultimately decided on this course of action to receive feedback from as many students as possible. Typically, only 10% of students give feedback for courses in our administrative system. With our method, we received feedback from 98 students (representing 94% of graded students).

Overall, students liked our concept of using ubiquitous platforms for dissemination of content with 83% favoring GitHub and YouTube over comparable self-hosted university services. We further asked detailed questions about the respective service usage below.

*YouTube usage.* We asked our students when they watched the videos on YouTube if 1) they noticed the timeline feature and 2) if they actively used it as shown in Figure 4. Almost all students generally watched the video – only 2 % did not watch them. Overall 93% noticed the timeline feature UI, as shown in Figure 2. The timeline feature has been actively used by 66 % of students.

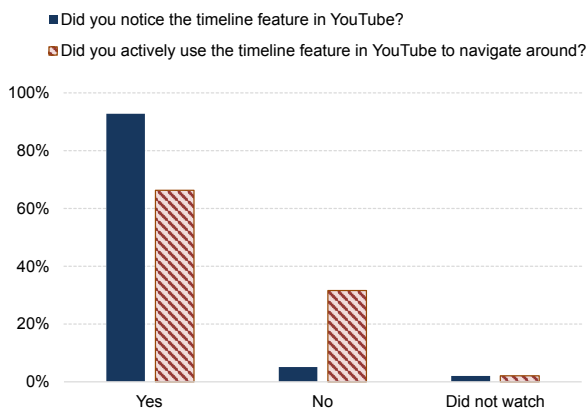

**Figure 4: Feedback result for YouTube usage of our timeline feature by our students.**

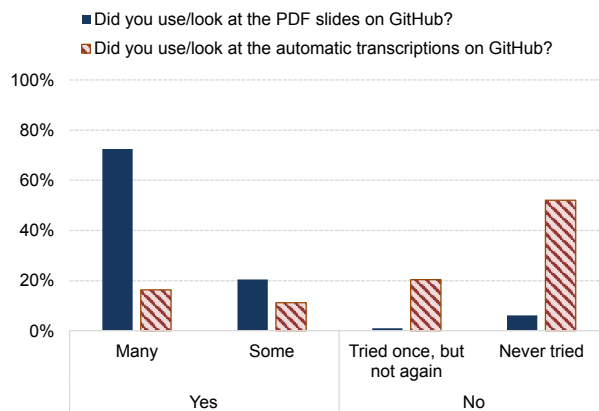

**Figure 5: Feedback result for GitHub usage of PDF slides and our transcriptions by our students.**

The high video watch response might draw suspicion of a bias in the feedback, albeit we did observe high view & unique user counts for the YouTube videos, giving us confidence that the feedback response is genuine.[9]

We see these results as a great success for adding timestamps to the videos. The UI is self-explanatory as most students noticed it and two thirds also actively used it. This is validated with written feedback that we got from our students, where many were very vocal about their surprise and the liking of this feature.

***GitHub usage.*** We asked students if they used the PDF slides or the transcripts: many times, sometimes, "only tried once, but not again", or never. We present the results in Figure 5.

We immediately see a striking difference between the two questions. The PDFs are viewed many times by 72% of students, and some of them still by 20%. Combined 92% of students regularly used the PDFs on GitHub. On the other hand our transcripts were only used many times by 16% and sometimes by 11% students. So roughly a quarter of our students used the transcripts. At first sight this looks like a bad outcome, however we have 52% of students that never tried to look at the transcripts.

For future iterations of this approach, we should explore changes in the way we incorporate the transcripts with the PDF slides. One such idea is to combine a single PDF slide with that exact transcript text, either in another PDF or a static HTML website. We plan to create an automated system that integrates into the GitHub workflow, whenever a new updated transcript is merged.

***Presence preferences.*** Now we look towards the future: We assume that at some point in the (near) future our university will move to in-person classes again. Because our remote approach was generally well received, we wanted to know how the availability of online materials would influence voluntary in-person attendance. In Figure 6 we compare the number of self-reported attendance frequency if the default only slide PDFs are online or additionally also video recordings available.

---

[9]The YouTube videos were always publicly available and therefore view counts not only include our students. Although, after the semester ended we observed a clear drop in new views, indicating that a majority of views came from our students.

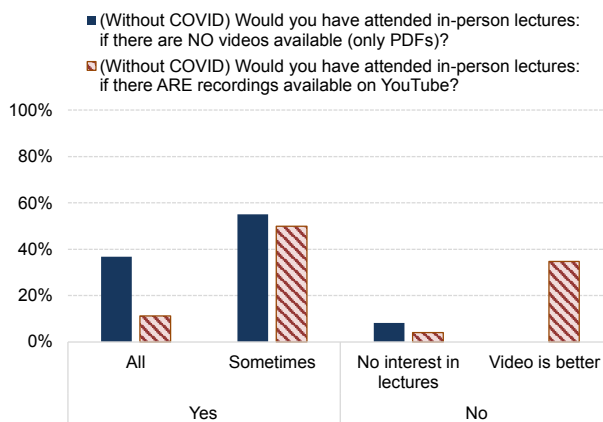

**Figure 6: Preferences on attending in-person lectures, once it is possible again, given available remote content.**

Most students are interested in attending some form of a lecture. Interestingly, the number of students that would attend all lectures drops substantially once recordings are available: from 37% to 11%. The number of students planning to attend some lectures stays roughly consistent around 50%. And finally, 35% of students say that they categorically like videos better and would not attend in-person lectures.

These results clearly indicate a benefit for a third of our students, if we have video recordings available. On the other hand this means that attendance will drop in our in-person lectures. Ultimately, we should offer both formats in a hybrid fashion and incentivize attendance in the classroom with more interactive settings in the future.

***Course quality.*** Following our workflow specific questions, we finally asked our students for an overall assessment of the course quality. In Figure 7 we present those results. We asked if students would recommend the lectures only or the whole course with lectures and exercises to a friend.

The lectures themselves are recommended by 98% of our students. Not a single student selected the "no way" answer. The whole course is recommended by 85%, where 13% would not really recommend it.

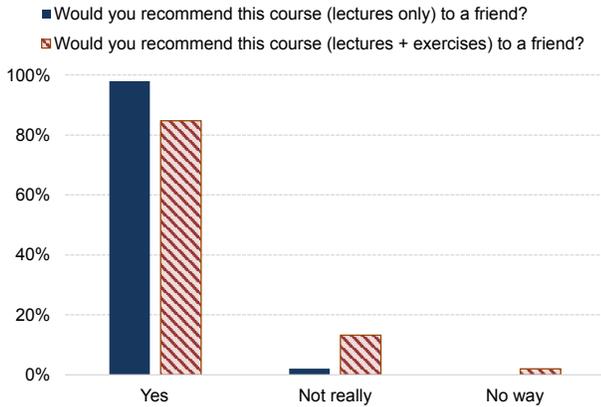

Figure 7: General course quality feedback, formulated as a question of recommending the course to a friend.

These numbers, of course, reflect our topic and content, however we argue that our didactic style with our content creation workflow is a major component of the students' satisfaction with our course. This also shows in the drop of recommendation between lectures and lectures+exercises, as we need to invest more time in better exercise descriptions.

*Take away.* Following the general positive feedback, we plan to repeat our general approach on the content creation and dissemination with the following planned improvements:

- We need to advertise the quality & readability of the transcripts more to our students, so that more students at least try to use them once.
- Combine slides and transcripts automatically in the student participation workflow in a usable format - such as a static website or PDF.
- Extract hyperlinks from the slide text and put them in the video description & transcript.

## 4 RELATED WORK

Similar to our work Angrave et al. [2] use prerecorded lecture videos plus transcriptions with a text-searchable interface in order to find relevant video moments. However, their solution requires hosting and maintaining a web-platform on-premise, which we did not want to do in order to reduce our workload.

Mason et al. [20] compare the effectiveness of an inverted classroom to a traditional classroom in a control-treatment experiment. In the inverted classroom, videos about the lecture material were posted on YouTube in conjunction with quizzes and homework assignments. While students initially struggled with the online format, they quickly adapted and evaluated the inverted classroom format satisfactory and effective. A similar picture arises in studies using online lectures with discussions [21] or lectures on YouTube [6]. They demonstrate that students find the online material extremely helpful and engage with it significantly more.

How the students interact with the online videos is explored by Guo et al. [12]. They show that shorter videos and informal talking-head videos are more engaging, similar to the findings of Amresh et al. [1]. Also Thompson and Petersen [22] explore high engagement with video content in an online learning environment and explore the interplay between the engagement with the video content and successfully answering questions about the videos content.

Besides YouTube, we emphasize the use of GitHub in order to enhance collaboration between the students as well as experience with version control systems. Britton and Berglund [5] reason the importance of experience using version control for students in order to work in industry and discuss how to early integrate learning version control in the study system. Haaranen and Lehtinen [13] evaluate the use of distributed version control systems for disseminating course material and show its success. Hsing and Gennarelli [15] compare the use of GitHub classroom to classrooms without GitHub and report better learning outcomes and classroom experiences. Similarly [4, 7, 11, 24] demonstrate the beneficial use of GitHub in online teaching organization. Malan et al. [19] develop an open-source tool which helps students with writing, testing and submitting programming assignments and helps teachers grade and check assignments for similarities. For students the tool includes feedback on errors in the code as well on coding style.

## 5 CONCLUSION

In this paper we presented our workflow for creating and disseminating lecture content for a remotely held university course. Our aim is to simultaneously fit the constraints of useful, enjoyable, and educational teaching materials while optimizing the time and resources spent on producing & publishing them. We provided a thorough description and code to replicate our efforts for any university course that includes online materials – be it completely remote or in a hybrid setting.

We record narrations for individual slides, which allows us to utilize our custom built export & transformation tool to effortlessly create transcripts and timestamps for the presentation. We utilize the timestamps in YouTube to let YouTube adapt the user interface of the video player to show slide changes. Furthermore, we utilize the collaborative GitHub platform to let our students voluntarily participate in improving the automatic transcripts.

Our students gave us overwhelmingly positive feedback: on the use of ubiquitous platforms, enabling YouTube's chapter UI through our automatically generated timestamps, as well continuing to use the automatic transcripts once they started. We experienced our workflow as truly time-saving and sustainable throughout a busy semester.

*Acknowledgements.* This work has received funding from the European Union's Horizon 2020 research and innovation program under grant agreement No 822670 and from the EU Horizon 2020 ITN/ETN project on Domain Specific Systems for Information Extraction and Retrieval (H2020-EU.1.3.1., ID: 860721).